\documentclass[aps,reprint,amsmath,amssymb]{revtex4-1}
\usepackage[utf8]{inputenc}
\usepackage{natbib}
\usepackage{color}
\usepackage{xcolor}
\usepackage{graphicx}
\usepackage{mdframed}
\usepackage{physics}
\usepackage{bm}
\patchcmd{\section}{\centering}{\raggedright}{}{}
\patchcmd{\subsection}{\centering}{\raggedright}{}{}
\begin{document}

\preprint{APS/123-QED}

\title{Testing the critical brain hypothesis using a phenomelogical renormalization group}

\author{Giorgio Nicoletti$^1$, Samir Suweis$^{1,2}$, Amos Maritan$^1$ \\}
\address{$^1$Dipartimento di Fisica ``G. Galilei'', Università di Padova \\
Via Marzolo 8, 35131 Padova, Italy \\
$^2$Padova Neuroscience Center \\ University of Padova, Padova, Italy
}

\begin{abstract}
\noindent \textit{We present a systematic study to test a recently introduced phenomenological renormalization group, proposed to coarse-grain data of neural activity from their correlation matrix. The approach allows, at least in principle, to establish whether the collective behavior of the network of spiking neurons is described by a non-Gaussian critical fixed point. We test this renormalization procedure in a variety of models focusing in particular on the contact process, which displays an absorbing phase transition at $\lambda = \lambda_c$ between a silent and an active state. We find that the results of the coarse-graining do not depend on the presence of long-range interactions, but some scaling features persist in the super-critical system up to a distance of $10\%$ from $\lambda_c$. Our results provide insights on the possible subtleties that one needs to consider when applying such phenomenological approaches directly to data to infer signatures of criticality.}
\\
\end{abstract}
\date{\today}

\maketitle

\noindent The possibility that living systems may be poised at criticality is a fascinating hypothesis \cite{bib:poised_criticality, bib:information_fitness, bib:colloquium}, and in recent years it has been explored in a vast variety of areas \cite{kinouchi2006optimal, bib:mora, bib:plenz, bib:flocks}.

Tools from statistical mechanics, such as the renormalization group \cite{bib:ma, bib:binney, bib:goldenfeld}, teach us that at criticality the macroscopic, collective behavior of the system is described by a few relevant attributes, such as the embedding dimension of the system and its symmetries, while most the microscopic details of the system become irrelevant. At the critical point the physical properties are determined by a non-trivial fixed point in the space of the possible models. However, in the broad landscape of natural systems one often has to deal directly with data without an explicit model, and the systems are typically finite, so that most of the time it is hard to come up with a definitive answer about whether they are poised near a critical point \cite{bib:colloquium}.

Recently, a phenomenological coarse-graining procedure was introduced in \cite{bib:bialek_prl, bib:cg_bialek} to deal with the long-range interactions that one reasonably expects in a network of neurons, but of which the full interaction network is not necessarily known. Data from single-neuron recordings, from the hippocampus of a mouse running along a virtual track, were directly analyzed by the authors in order to understand if this coarse-graining procedure (which we recall in Section 1) drives the system towards a non-trivial fixed point in the renormalization group sense, hence if the neural dynamics is critical and details independent. Indeed, the brain is probably one of the most impressively complex system we are able to study and the idea that the collective behavior of neurons might emerge from a self-organized critical state has been widely studied in the last year \cite{bib:chialvo, bib:thermodynamic_neurons, bib:information_fitness,bib:rocha2018homeostatic}.

One of the first evidences that suggested this hypothesis is the presence of neuronal avalanches that spontaneously occurs in the brain, i.e. during spontaneous activity, that show a spatio-temporal power law distribution with exponents compatible with those of a mean field branching process \cite{bib:plenz}. However, this conclusion is highly debated, in fact, and the same exponents may stem, for instance, from an underlying non-critical neutral \cite{bib:neutral} or random \cite{bib:touboul2017power} dynamics and, in general, the subject is far from being settled. 

In this paper, we aim to test this phenomenological renormalization group (PRG) method by applying it to a well known non-equilibrium statistical model, the contact process \cite{bib:harris}. This model belongs to the universality class of directed percolation and displays an absorbing phase transition, which has been widely studied \cite{bib:dickman}. Its critical behavior is well understood and the exponents are known from numerical studies, so we shall regard it as a ``control case'' to investigate the ability of this procedure to extract the relevant information and infer signatures of a critical state in out-of-equilibrium systems.

On a $d$-dimensional hyper-cubic lattice with nearest neighbor interactions it is sufficient to introduce long-range connections to change the topology, for instance, to that of a small-world network. Hence by simulating the contact process we are able in particular to probe the impact of short and long range interactions on the coarse grained system behaviour, and in particular we are able to shed some light on the possible outcomes and interpretations of the emergent fixed point describing the system collective behavior. Along the road, we also test the PRG procedure in other models to better characterize its results.

\section{The coarse-graining procedure}
\label{coarse_graining}
\noindent In this section we briefly describe the coarse-graining procedure introduced in \cite{bib:cg_bialek, bib:bialek_prl} that we aim to test. The authors propose to build clusters of variables by grouping together neurons that are most correlated, so that the overall correlation structure tends do be preserved.

Let us consider a system (e.g. neural circuit) of $N$ variables (e.g. neurons) connected by a given, but typically unknown, interaction network. Denoting state variables of the neurons  as $\sigma_i^{(1)}$ for $i=1,...,N$, where the superscript 1 denotes that we are at the first step of the renormalization procedure, we search for the maximal non-diagonal element of the normalized correlation matrix
\begin{align*}
c_{ij} = \frac{C_{ij}}{\sqrt{C_{ii}C_{jj}}}
\end{align*}
where $C_{ij}$ is the covariance matrix
\begin{align*}
C_{ij} = \ev{\sigma_i^{(1)}\sigma_j^{(1)}} - \ev{\sigma_i^{(1)}}\ev{\sigma_j^{(1)}},
\end{align*}
where $\langle\cdot\rangle$ represents the average over the time-series of neural activity. The pair $(i, j_*(i))$ of maximally correlated variables is removed and we search again, ending up with a set of pairs $\{i, j_*(i)\}$. The coarse-grained variables are defined as
\begin{align*}
\sigma_i^{(2)} = \sigma_i^{(1)} + \sigma_{j_*(i)}^{(1)} 
\end{align*}
where $i = 1, \dots, N/2$. We iterate this process, producing clusters of $K=1, 2, 4, \dots, 2^{k-1}$ variables. Each one defines a new variable $\sigma_i^{(k)}$ as the summed activity of cluster $i$.

Under this coarse-graining procedure, in \cite{bib:cg_bialek, bib:bialek_prl} the behavior of various quantities is analyzed in order to make some parallels with the behavior of critical systems. In particular, the following observables are studied: the mean variance of the neural activity; the distribution of the individual coarse-grained variables; the spectrum of the covariance matrix; and the mean autocorrelation function.

The mean variance of the activity is defined as
\begin{align}
\label{eqn:cg_variance}
M_2(K) = \frac{1}{N_k} \sum_{i=1}^{N_k} \left[\left\langle \left(\sigma_i^{(k)}\right)^2 \right\rangle - \left\langle\sigma_i^{(k)}\right\rangle^2\right]
\end{align}
where $N_k$ is the number of variables after $k$ steps of the coarse-graining procedure. If the variables are independent one would obtain a variance scaling as $M_2(K) \propto K^{\tilde\alpha}$ with $\tilde\alpha = 1$.

More generally, we can study the full distribution of the individual coarse-grained variables. Since a coarse-grained variable $\sigma_i^{(k)}$ is vanishing if and only if all the $2^{k-1}$ raw variables are zero \footnote{In fact, we assume that $\sigma_i \ge 0$, $\forall i$}, we can write
\begin{align*}
P\left(\sigma_i^{(k)}\right) = & \, P_\text{silence}(K) \delta\left(\sigma_i^{(k)}, 0\right) \\
& + \left[ 1 - P_\text{silence}(K)\right]A_K\left(\sigma_i^{(k)}/K\right)
\end{align*}
where $K=2^{k-1}$ and
\begin{align}
\label{eqn:full_probability}
P_\text{activity}(\sigma_i^{(k)}/K) = A_K\left(\sigma_i^{(k)}/K\right)
\end{align}
is the probability distribution of the normalized activity. Thus we look at an effective (reduced) free energy
\begin{align}
\label{eqn:cg_free_energy}
F(K) = \log P_\text{silence}
\end{align}
(this formula is based on the assumption that the {\it energy} of the system is zero when no activity is present) and at its possible scaling $F \sim -K^{\tilde \beta}$. For independent variables, we expect $\tilde\beta = 1$.

A scaling behavior of the ranked spectrum of the covariance matrix at the critical point is expected. In fact, the correlation function decays algebraically as $G(\mathbf x) \sim |\mathbf x|^{-(d-2+\eta)}$, and one can show (see Appendix A) that in translational invariant systems the eigenvalues of the covariance matrix scale as
\begin{align*}
\lambda_{r} \sim r^{-(2-\eta)/d}.
\end{align*}
where $r$ is the rank of $\lambda_r$, ordered from the highest to the smallest. If we consider the variables inside the clusters that we build along the coarse-graining, the highest possible rank $r$ is given by the number of variables $K$ that make up each cluster. Hence at criticality we should find
\begin{align}
\label{eqn:cg_eig}
\lambda_r \propto \left(\frac{K}{r}\right)^\mu
\end{align}
with $\mu = (2-\eta)/d$, and this is a direct consequence of the power law decay of the correlation function in space.

Finally, the mean autocorrelation function is obtained by
\begin{align}
\label{eqn:cg_acf}
C^{(k)}(t) = \frac{1}{N_k}\sum_i C_i^{(k)}(t)
\end{align}
where
\begin{align*}
C_i^{(k)}(t) = \frac{\langle \sigma_i^{(k)}(t_0)\sigma_i^{(k)}(t_0+t)\rangle - \langle \sigma_i^{(k)}\rangle^2}{\langle(\sigma_i^{(k)})^2\rangle - \langle \sigma_i^{(k)}\rangle^2}.
\end{align*}
Since we are grouping correlated variables, the decay of the autocorrelation is slower in clusters of bigger size. However, in a critical system we might expect dynamical scaling, which would imply a power law scaling of the autocorrelation times $\tau_c \propto K^{\tilde z}$.

\begin{figure}[t]
\centering

\begin{minipage}{0.2\textwidth}
        \centering
        \includegraphics[height=3.5cm]{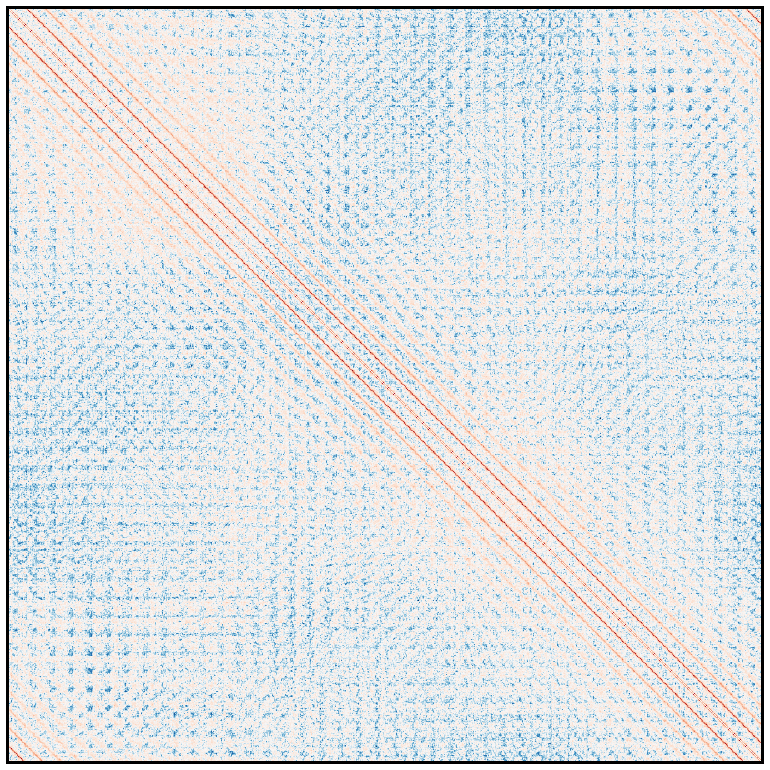}
\end{minipage}
\hspace{0.2cm}
\begin{minipage}{0.23\textwidth}
        \centering
        \includegraphics[height=3.5cm]{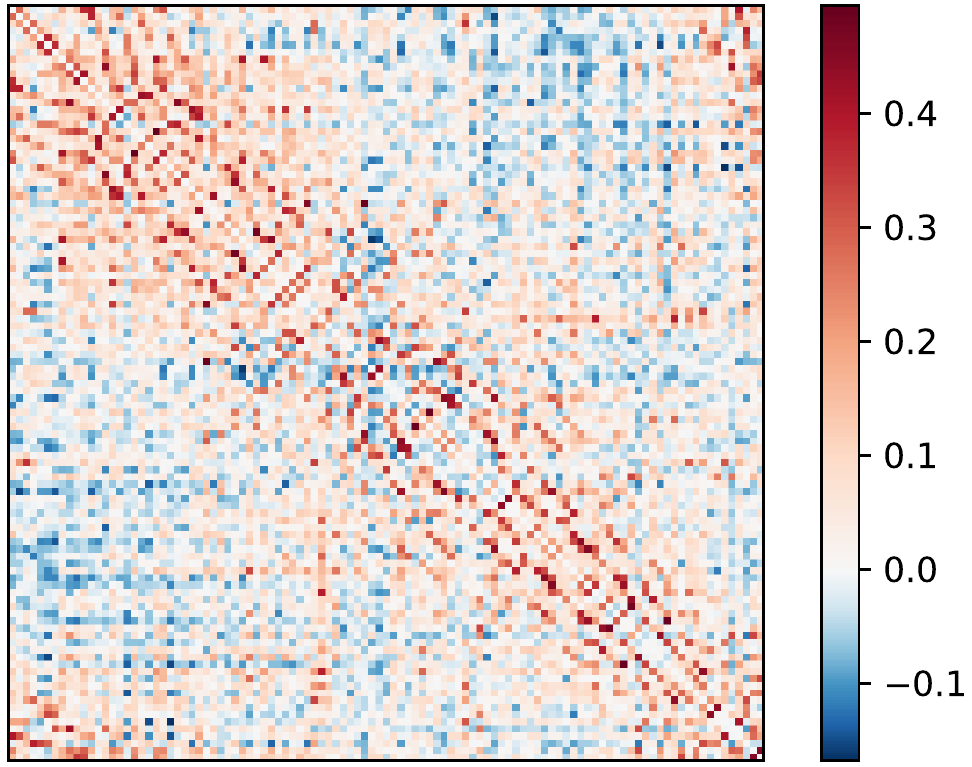}
\end{minipage}
\caption{\footnotesize The correlation matrix for the $2D$ contact process at $\lambda = \lambda_c$. Left panel: correlation matrix of the raw variables. Right panel: correlation matrix of clusters of $16$ variables. Notice how the coarse-graining seems to preserve and unravel the non-trivial correlation structure.}
    \label{fig:corr_mat}
\end{figure}

A different test can be performed by exploiting the fact that, in systems with translational invariance, the eigenvalues $\lambda_{\vb k}$ of the covariance matrix in momentum space are the Fourier transform of the correlation function $G(\vb k)$. Since coarse-graining in momentum space amounts to average over the  Fourier modes with small wavelength, we expect that averaging over low variance contributions in the covariance matrix should lead to an equivalent result. Hence, we consider the set of eigenvectors of the covariance matrix $\{\vb{u}_r\}$, ordered according to the value of the corresponding eigenvalue, from the the highest to the smallest one, and we introduce the projectors
\begin{align}
\label{eqn:projector}
P_{ij}(K) = \sum_{r=1}^K u_{ir}u_{jr}
\end{align}
where $P_{ij}(N)$ is the identity, hence the eigenvectors are orthonormalized. The authors of \cite{bib:cg_bialek, bib:bialek_prl} propose to consider a cutoff $\hat{K}<N$, in analogy to the cutoff in momentum space, in such a way that the low variance contributions do not enter the projector (\ref{eqn:projector}. Then the coarse-grained variables are defined as
\begin{align}
\phi_i(\hat K) = z_i(\hat K) \sum_j P_{ij}(\hat K) \left[\sigma_j^{(1)}-\langle \sigma_j^{(1)}\rangle\right]
\end{align}
where $z_i(\hat K)$ assures that the coarse-grained variables have unitary variance, i.e., $\langle \phi^2_i(\hat K) \rangle = 1$. By means of the Young-Eckart theorem \cite{bib:young_eckart}, the above procedure allows one to find the best decomposition with rank $\hat K$ of the original data matrix. In this setting it is interesting to look at the distribution
\begin{align}
\label{eqn:cg_momentum_dist}
P_{\hat K}(\phi) = \big\langle \frac{1}{N} \sum_{i=1}^N \delta\left(\phi_i(\hat K) - \phi\right)\big\rangle=\frac{1}{N} \sum_{i=1}^N \mathbb{P}\left[\phi_i(\hat K) = \phi\right]
\end{align}
as we change the cutoff $\hat K$.
In fact, a renormalization group transformation typically drives the joint probability towards a fixed point, and if the variables are weakly correlated such fixed point is the one obtained from the central limit theorem \cite{bib:rg_probability}. Hence, the authors of \cite{bib:cg_bialek, bib:bialek_prl} propose to use this PRG approach to test whether the joint distribution converges towards a non-Gaussian critical fixed point.


\section{The model}
\label{contact_process}
\noindent The contact process \cite{bib:harris, bib:dickman, bib:henkel} is possibly the simplest non-equilibrium model used to describe the propagation of neural activity on a network \cite{bib:neutral, bib:rocha2018homeostatic}. Hence, it is a proper modelling framework to test the coarse-graining procedure just described.

Consider a collection of $N$ nodes of a given network. Each node can be either active (occupied) or inactive (empty), and we identify its state by means of a binary variable $\sigma_i (t) = 1, \, 0$ respectively. The activity spreads via a nearest neighbors interaction, and it depends on the number of active neighbors $n_i(t) = \sum_{j\in\ev{i}} \sigma_j(t)$, whereas each active site is emptied at a unitary rate. The rates $w[\sigma_i(t)\to\sigma_i(t+dt)| n_i(t)]$ that define the process for a node with $k_i$ neighbors are given by
\begin{equation}
\label{eqn:cp_rates}
w[0\to1| n_i] = \frac{\lambda n_i}{k_i}, \quad 
w[1\to0| n_i] = 1
\end{equation}
where $\lambda$ is the spreading rate.

The configuration with all empty sites is an absorbing state, since the system cannot escape from it. In particular, if $\lambda>\lambda_c$ the stationary state is an active fluctuating phase, whereas if $\lambda<\lambda_c$ the system eventually gets trapped in the absorbing configuration. Hence, we choose the density $\rho$ of the active sites as an order parameter, while $\lambda$ is the control parameter. In fact, exactly at $\lambda = \lambda_c$ the density of active sites undergoes large fluctuations and the system is often close to the absorbing state \footnote{Indeed one can prove that the critical contact process dies out with probability $1$ \cite{bib:cp_diesout}. A non vanishing survival probability is achieved only in the super-critical regime $\lambda>\lambda_c$.}.

The nature of the phase diagram can be readily understood from the mean field approximation
\begin{equation}
\label{eqn:cp_mf}
\dot{\rho} = \rho(\lambda-1)-\lambda\rho^2.
\end{equation}
This equation has two stationary solutions:  $\rho_\text{st}^v = 0$ and the active state $\rho_\text{st}^a = (\lambda-1)/\lambda$. The former is stable if $\lambda < 1$, and the latter if $\lambda >1$. Hence the mean field critical point is $\lambda_{c}^\text{MF} = 1$.

The contact process is not exactly solvable even in one dimension, therefore we need to rely on numerical studies. We implement the usual scheme \cite{bib:sim_cp}: an occupied site $i$ is randomly chosen, and with probability $1-p_\lambda = 1/(1+\lambda)$ the site is emptied. With probability $p_\lambda = \lambda/(1+\lambda)$ one of the neighbors is picked at random and, if empty, is occupied. The time is increased by $1/N_\text{occ}$, where $N_\text{occ}$ is the number of occupied sites.

We are interested in two different types of interaction network topology to test their effect on the coarse-graining procedure: a $2D$ lattice with periodic boundary conditions and a small-world network. The former is a rather standard choice, and the estimated critical point is $\lambda_c^{2D} \approx 1.6488$ \cite{bib:dickman}. On the other hand, the latter setting is more realistic, given the existence of long synaptic connections occurring in a network of neurons.

In general, it would be ideal to have a coarse-graining procedure that works both for short-range and long-range interactions, especially if one needs to deal directly with neural activity data and the specific network architecture is not accessible. In the small-world case the critical point $\lambda_c^\text{SW}$ depends on the rewiring probability, and it has been studied numerically in \cite{bib:cp_smallworld}.

We perform all the simulation with $N = 40^2$ sites and analyze clusters of size $K = 2, \dots, 256$. In momentum space, we keep up to $N/128 \approx 12 $ eigenvalues, which is less than $1\%$ of the original modes.

\begin{figure}[!t]
\centering
\begin{minipage}{0.23\textwidth}
        \centering
        \includegraphics[height=4cm]{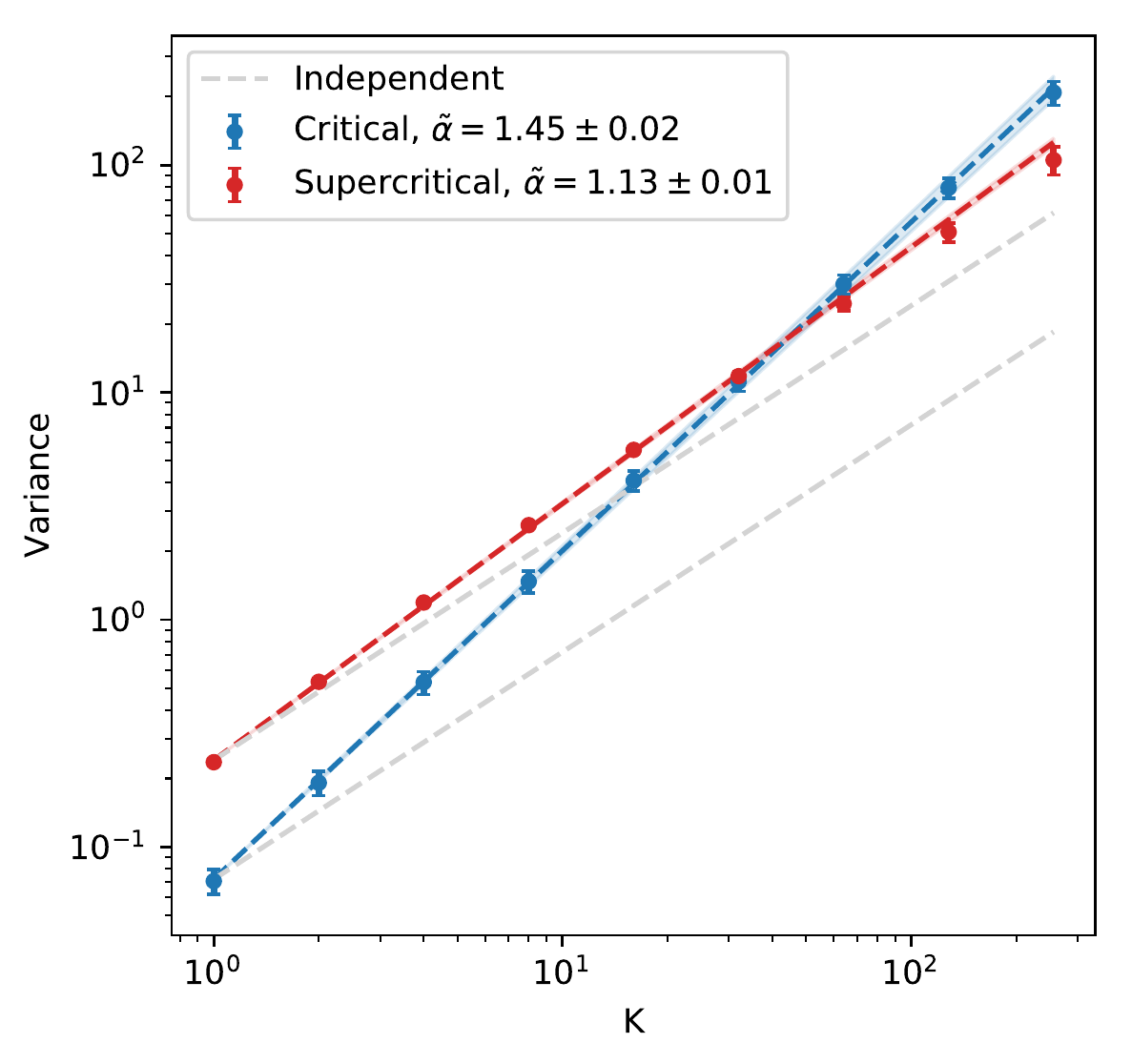}
    \end{minipage}
    \hfill
    \begin{minipage}{0.23\textwidth}
        \centering
        \includegraphics[height=4cm]{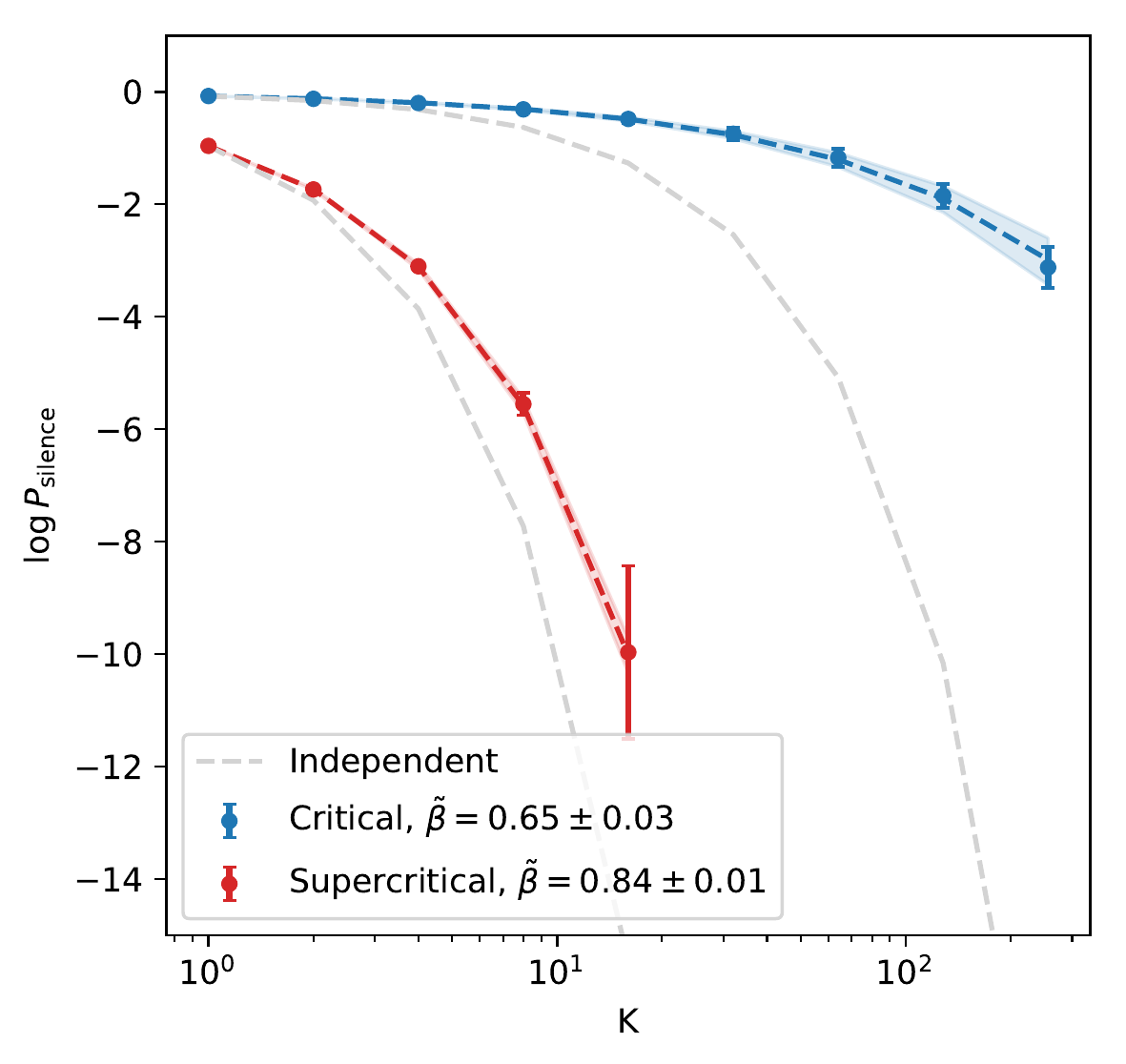}
    \end{minipage}
    \caption{\footnotesize {Left panel: scaling of the variance, Equation (\ref{eqn:cg_variance}). Right panel: scaling of the free energy, Equation (\ref{eqn:cg_free_energy}). Both are fitted with the corresponding power laws shown in Section 1. Notice that both the critical and the super-critical regime show power law behaviour but with different exponents: in particular, the silence probability is smaller and decays much faster in the super-critical case due to the proliferation of the activity. However, we do not find full compatibility between the super-critical contact process and the independent case.}}
    \label{fig:var_sp}
\end{figure}

\section{Results}
\begin{figure}[t]
\centering
\begin{minipage}{0.23\textwidth}
        \centering
        \includegraphics[height=4cm]{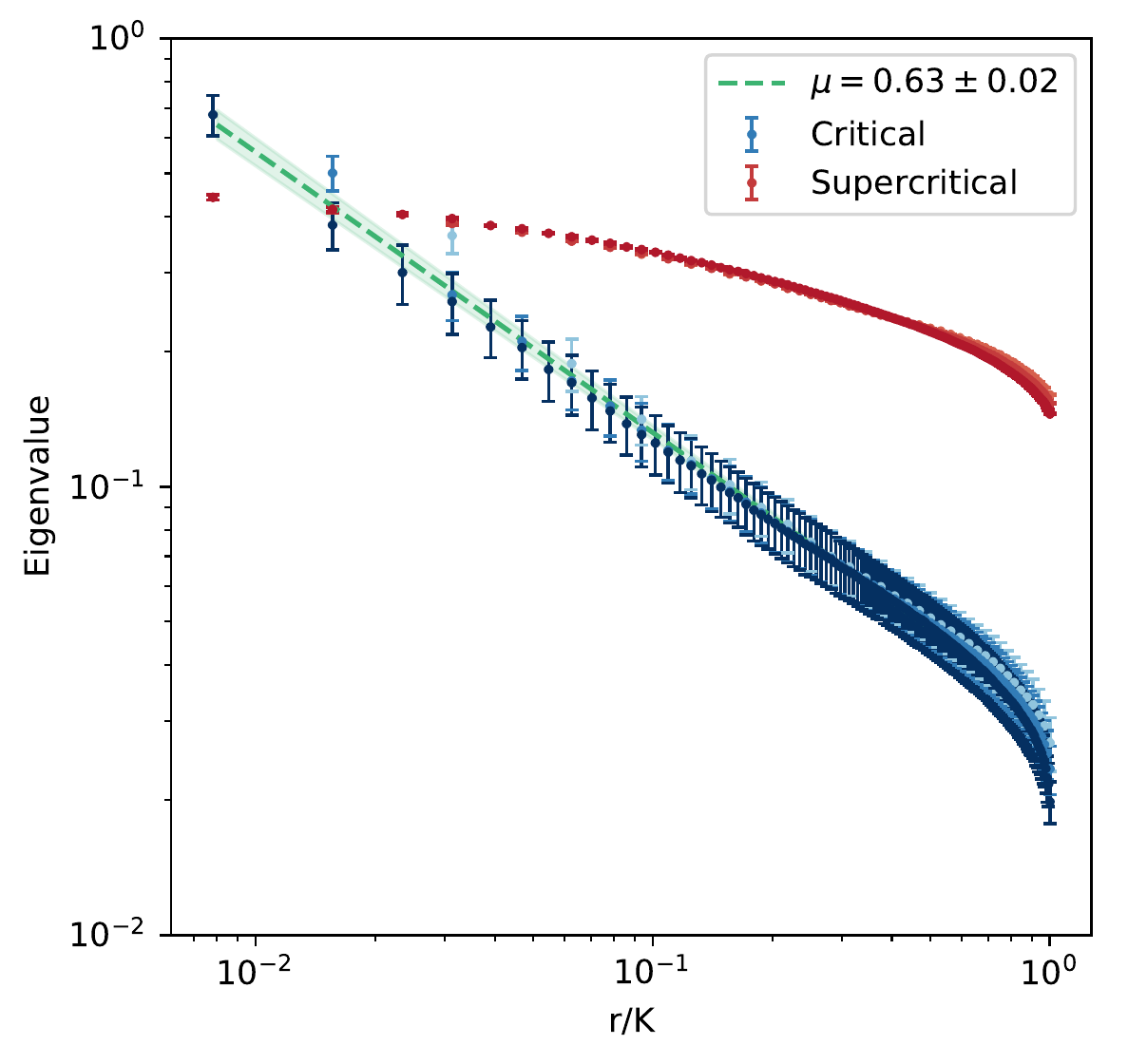}
    \end{minipage}
    \hfill
    \begin{minipage}{0.23\textwidth}
        \centering
        \includegraphics[height=4cm]{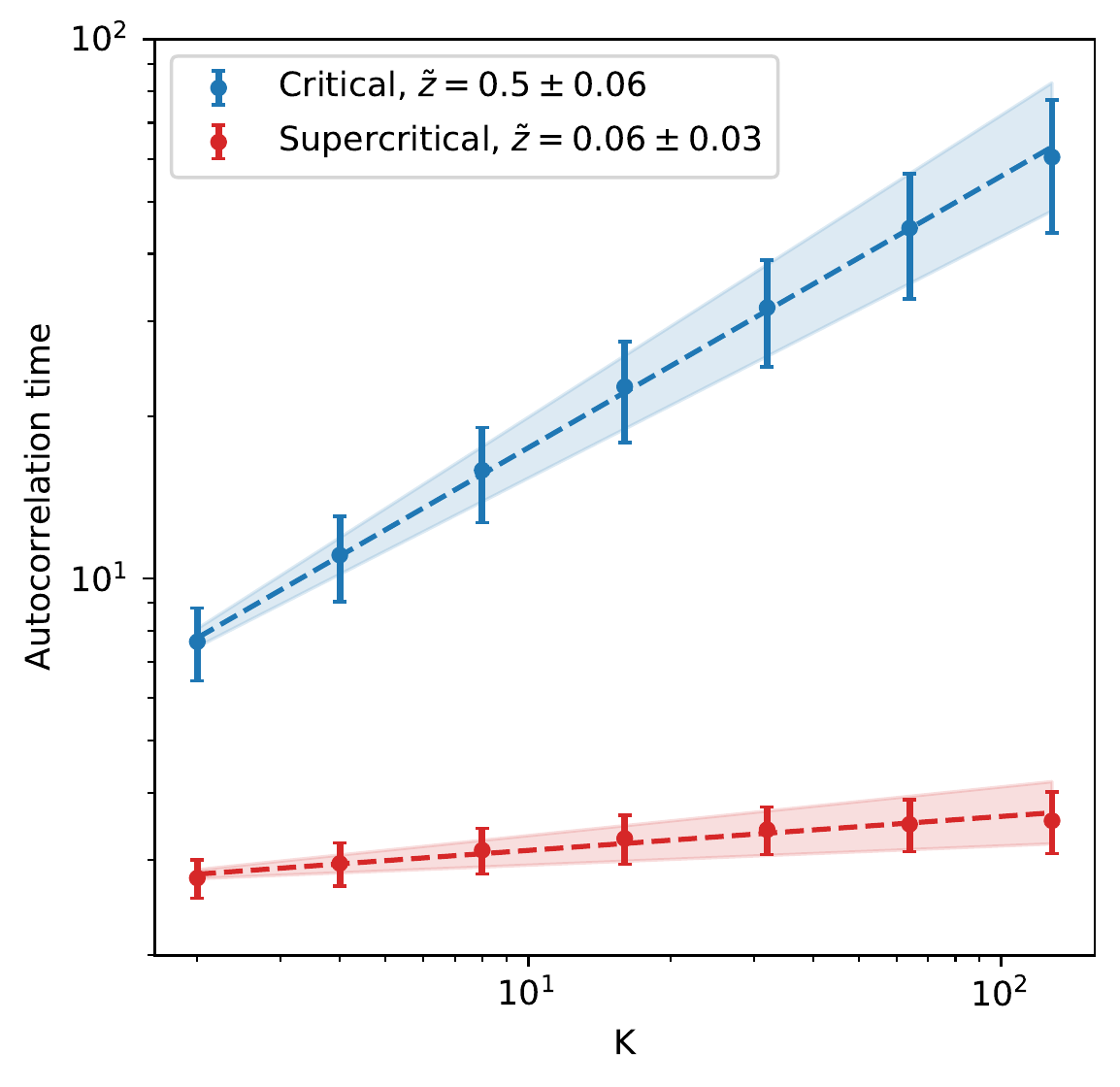}
    \end{minipage}
    \caption{\footnotesize Left panel: scaling of the eigenvalues of the covariance matrix inside clusters of size $K = 32, 64, 128$. Right panel: scaling of the autocorrelation times during coarse-graining. The distinction among the two phases is rather clear. For the spectrum, in the critical case we find an exponent $\mu = 0.63\pm0.02$ compatible with the expected value, whereas the eigenvalues show less variability at $\lambda > \lambda_c$. The same holds for the autocorrelation times, which follow a power law behavior at criticality and become negligible above it.}
    \label{fig:corr_structure}
\end{figure}

\noindent We now consider the contact process in a $2D$ lattice, both at $\lambda = \lambda_c^{2D}\approx 1.6488$ and in the super-critical phase at $\lambda = 3$. We will show that the PRG is indeed able to distinguish between these two different phases as the coarse-graining procedure gives indeed results that are rather different. However, we find the existence of some caveats that is important to take into account.

In Figure \ref{fig:var_sp} we see that in the super-critical regime the exponents of the variance and of the free energy are not exactly compatible to the independent case $\tilde \alpha = 1 = \tilde\beta$. Nevertheless, the profile of the free energy clearly shows that the underlying dynamics is different, as for clusters of size $K>32$ the silence probability vanishes in the active phase. Notably, if we compare the critical exponent $\tilde\beta = 0.65 \pm 0.02$ with the one obtained for real neurons in \cite{bib:cg_bialek, bib:bialek_prl}, $\tilde\beta_\text{neurons} = 0.893 \pm 0.003$ we see that in the contact process the decay of the silence probability with the cluster size is slightly slower, meaning that the sites tend to be more active than real neurons.

\begin{figure}[t]
\centering
\begin{minipage}{0.23\textwidth}
        \centering
        \includegraphics[height=4cm]{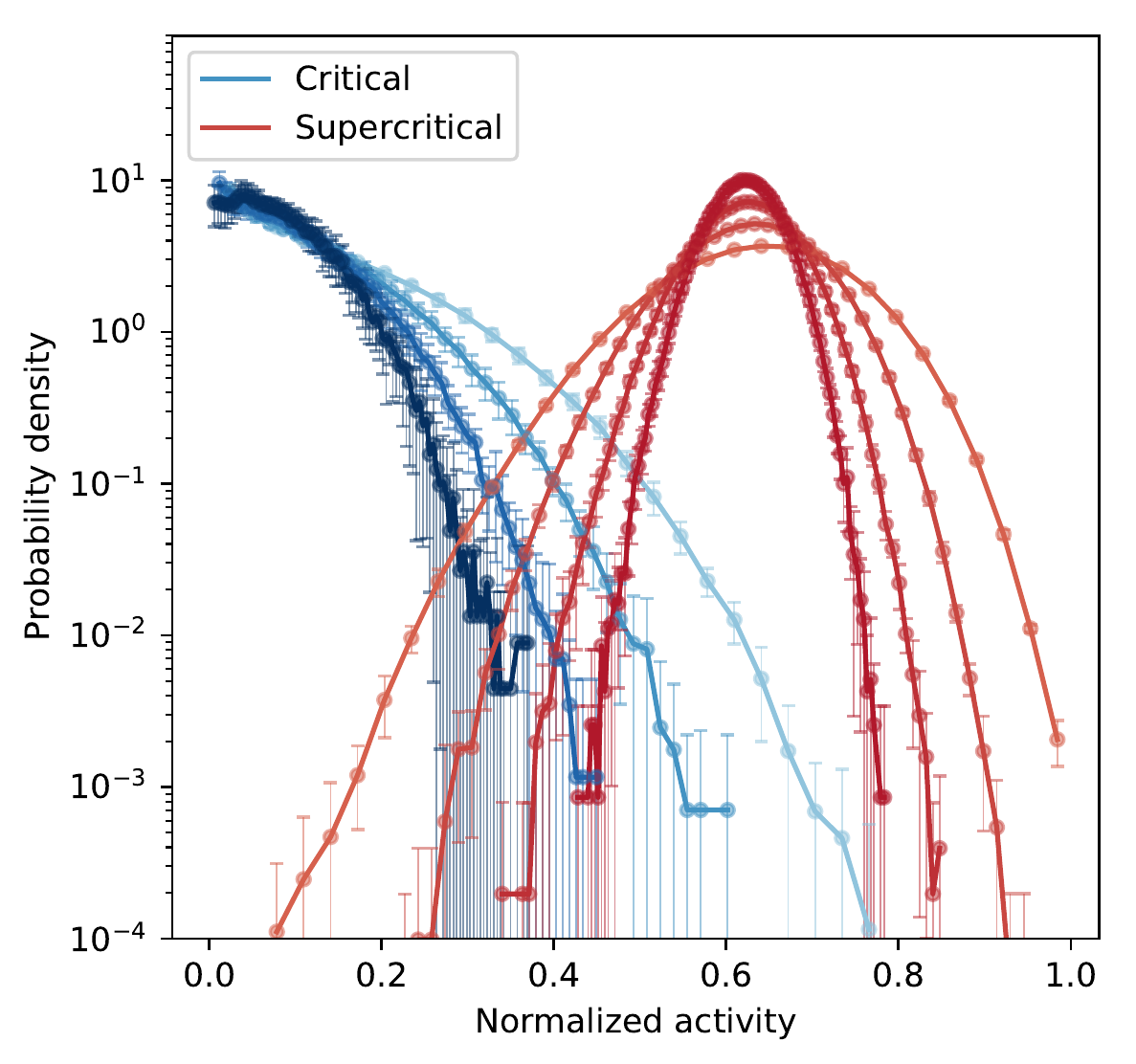}
    \end{minipage}
    \hfill
    \begin{minipage}{0.23\textwidth}
        \centering
        \includegraphics[height=4cm]{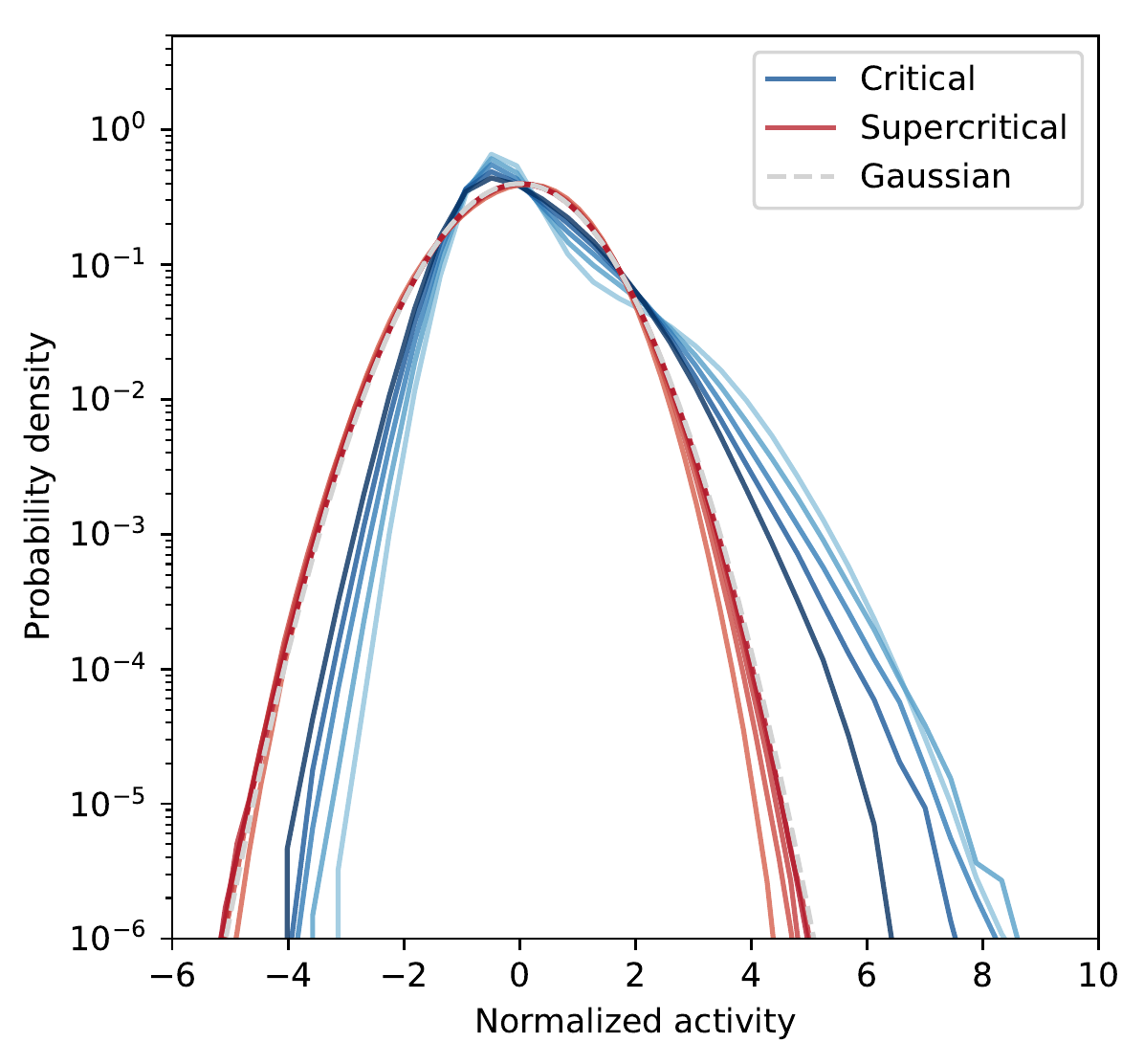}
    \end{minipage}
    \caption{\footnotesize Left panel: evolution of the probability distribution of the non-zero normalized activity during the coarse-graining via maximally correlated pairs, as in Equation (\ref{eqn:full_probability}). Right panel: evolution of the probability distribution of the coarse-grained variables in momentum space, as in Equation (\ref{eqn:cg_momentum_dist}). We show the results for $K = 32, \dots, 256$ and for $N/8, \dots, N/128$ modes (from brighter to darker color). The distribution in direct space is very different due to the critical contact process being typically close to the absorbing state. In momentum space, the super-critical contact process converges to a Gaussian fixed point in agreement with to the central limit theorem \cite{bib:rg_probability}, whereas at criticality we see the presence of non-Gaussian tails.}
    \label{fig:probability}
\end{figure}

Figure \ref{fig:corr_structure} shows the correlation structure of the system's quasi-stationary state. The change in the spectrum of the covariance matrix is more evident, since in the super-critical case the eigenvalues span a smaller set of values. In the critical case, instead, we find a power law decay with an exponent $\mu = 0.63 \pm 0.02$, with $\mu = (2-\eta)/d$. In real neurons, \cite{bib:bialek_prl, bib:cg_bialek} report $\mu_\text{neurons} = = 0.71 \pm 0.06$. We note that since one of the hyper-scaling relations of the contact process yields \cite{bib:qs_corr}
\begin{align*}
\eta = d -2 + \frac{\beta}{\nu_\perp},
\end{align*}
we expect $\mu \approx 0.6$, which is compatible with what we find using the PRG procedure. The time-autocorrelation function shows an evident change as well: in the super-critical regime the autocorrelation decays exponentially, whereas at criticality we find a power scaling with an exponent $\tilde z = 0.50 \pm 0.06$. We note also that in the super-critical regime a power-law seems to be present, but the small exponent is compatible with the absence of scaling \footnote{A constant autocorrelation time across different cluster sizes fits with the same significance the data, hence no relevant scaling feature seems to be present in the supercritical regime.}.

The evolution of the joint probability distribution of the coarse-grained variables in Figure \ref{fig:probability} shows once more the differences in the underlying dynamics. The most notable result is the convergence in momentum space: for $\lambda > \lambda_c^{2D}$ the fixed point is Gaussian in accord with the central limit theorem, whereas at $\lambda = \lambda_c^{2D}$ we see distinct non-Gaussian tails. The last coarse-graining step in momentum space only keeps $N/128$ modes, so the fact that we still see non-trivial tails is significant.

\begin{figure}[t]
\centering
\begin{minipage}{0.23\textwidth}
        \centering
        \includegraphics[height=4cm]{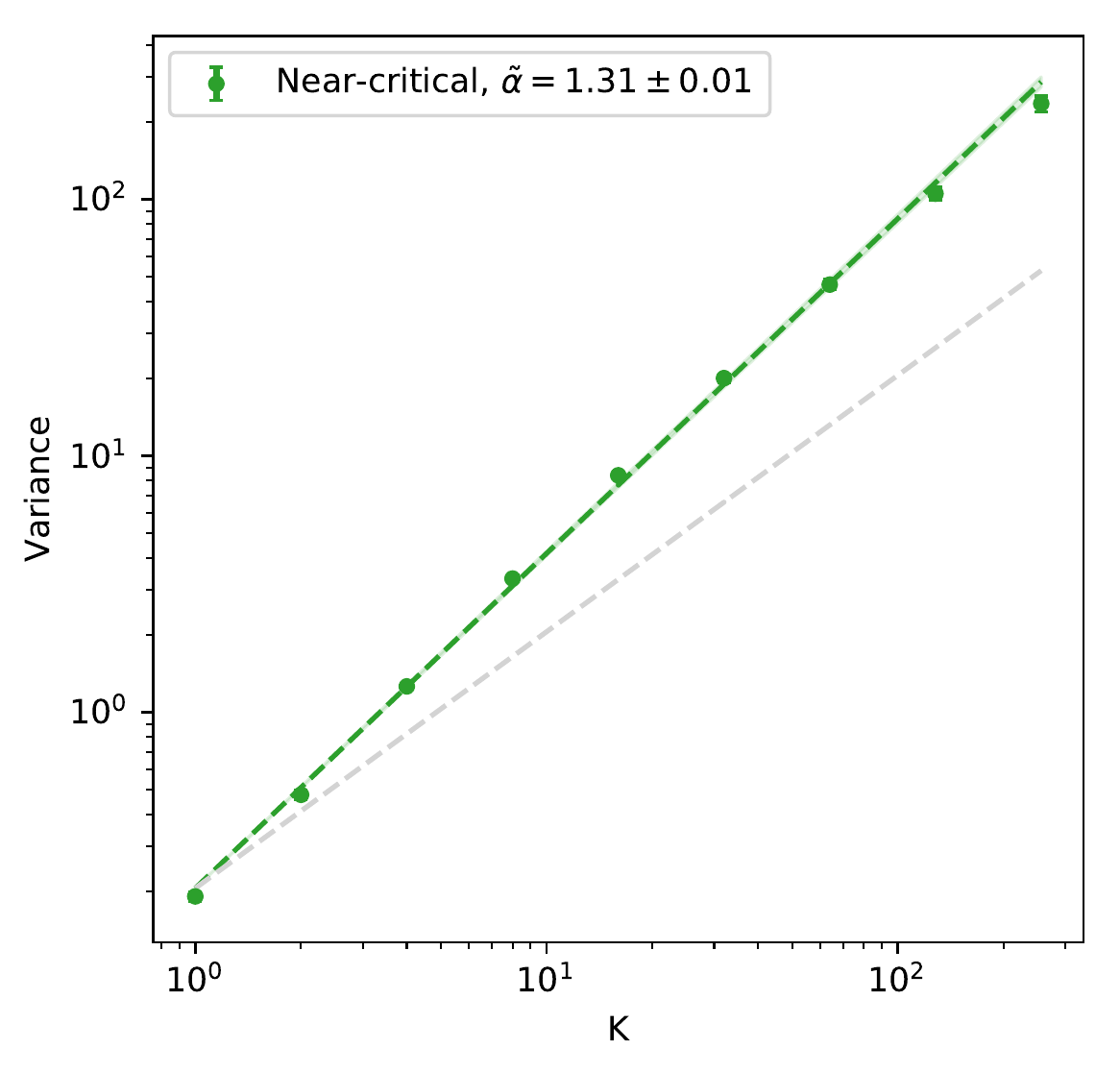}
    \end{minipage}
    \hfill
    \begin{minipage}{0.23\textwidth}
        \centering
        \includegraphics[height=4cm]{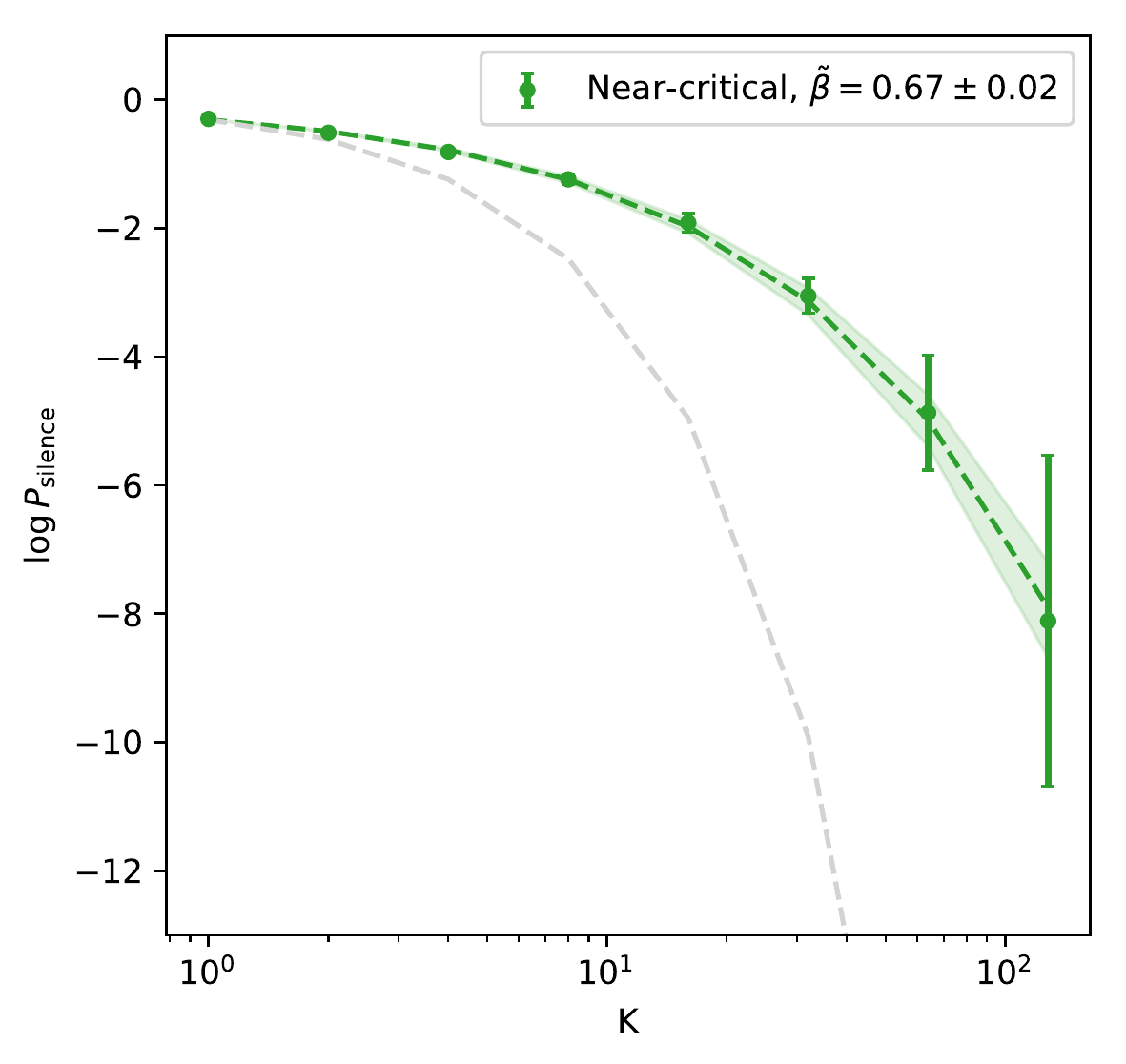}
    \end{minipage}
    \caption{\footnotesize Left panel: scaling of the variance (\ref{eqn:cg_variance}) during the coarse-graining for $\lambda = \lambda_\text{nc}$. Right panel: scaling of the free energy (\ref{eqn:cg_free_energy}) at $\lambda = \lambda_\text{nc}$. Even if this is a super-critical contact process, the exponents are far from being comparable with the independent case $\tilde\alpha = 1 = \tilde\beta$. Overall, the scaling is more similar to the real critical case, see Figure \ref{fig:var_sp}. Notice in particular how the silence probability is non vanishing even for large clusters.}
    \label{fig:nc_var_sp}
\end{figure}

These results prove to be very stable when we change the underlying topology and we introduce long-range connections by choosing a small-world network. In particular, we implement a Watts-Strogatz model \cite{bib:smallworld} with a rewiring probability $p = 0.01$. The critical point is $\lambda_c^\text{SW} \approx 1.7961$, as given by \cite{bib:cp_smallworld}. We find that all the considerations we made so far hold in the small-world topology as well, and the presence of long-range interactions does not affect the results of the PRG coarse-graining procedure.

As a sanity check, we also implement a synchronous update algorithm. The results show no difference with respect to the asynchronous one used insofar \footnote{In the case of the asynchronous update, where at most one site is changed at each step, one should carefully consider that the algorithm induces a spurious correlation between subsequent configurations. Hence we do not keep all the configurations to perform averages but we sub-sample them so to select only uncorrelated configurations. Once this is taken in account, the results of synchronous and asynchronous updates are equivalent}.

\subsection{Persistence of the scaling near a critical point}
\noindent A natural question one may ask is how sensible this PRG approach is, i.e., how easy it is to distinguish a truly critical system from a super-critical one. We test this in the contact process by moving the control parameter from the critical point $\lambda_c$ to $\lambda_\text{nc} \approx 1.1\lambda_c$, that is a $10\%$ increase. Notice that, although it is not trivial to define a finite-size critical point \cite{bib:finitesize_ising} for the transition in the contact process \footnote{Indeed, the finite size contact process eventually reaches the absorbing configuration, at all $\lambda$ \cite{bib:dickman}}, at $\lambda_\text{nc}$ we do see distinctive features of a super-critical dynamics: in fact, we do not see considerable fluctuations in the density of sites, nor the system constantly approaches the absorbing state as at $\lambda = \lambda_c$. Hence, at $\lambda_\text{nc}$ the dynamical evolution is significantly super-critical, and we shall refer to this as a near-critical case to distinguish it from the super-critical regime we described before.

In Figure \ref{fig:nc_var_sp} we see non-trivial scaling behaviors of both the variance and the free energy. If we compare them to Figure \ref{fig:var_sp}, they are arguably more similar to the critical regime rather than the super-critical one. Noticeably, the exponents $\tilde{\alpha}$ and $\tilde{\beta}$ are in between the two cases (i.e., critical and super-critical ones), suggesting that as $\lambda$ smoothly changes from $\lambda = \lambda_c$ to $\lambda = +\infty$, the exponents smoothly approach $1$. This result calls for carefulness as the scaling inferred from the PRG of the variance and of the free energy are not related only to the system critical state \footnote{The fact that both the variance (\ref{eqn:cg_variance}) and the free energy defined in (\ref{eqn:cg_free_energy}) show a power law behavior both at $lambda_c$ and at $\lambda > \lambda_c$ might be a sign that criticality is not a necessary condition for these quantities to scale as a power law. In particular, in the case of the free energy one should note that usually it is the singular part of the free energy that shows scaling, whereas with this PRG we cannot distinguish it from the non-singular part.}. Indeed, the difficulty to distinguish between critical or quasi critical states is confirmed also from other studies (using different approaches) \cite{priesemann2014spike}.

\begin{figure}[t]
\centering
\begin{minipage}{0.23\textwidth}
        \centering
        \includegraphics[height=4cm]{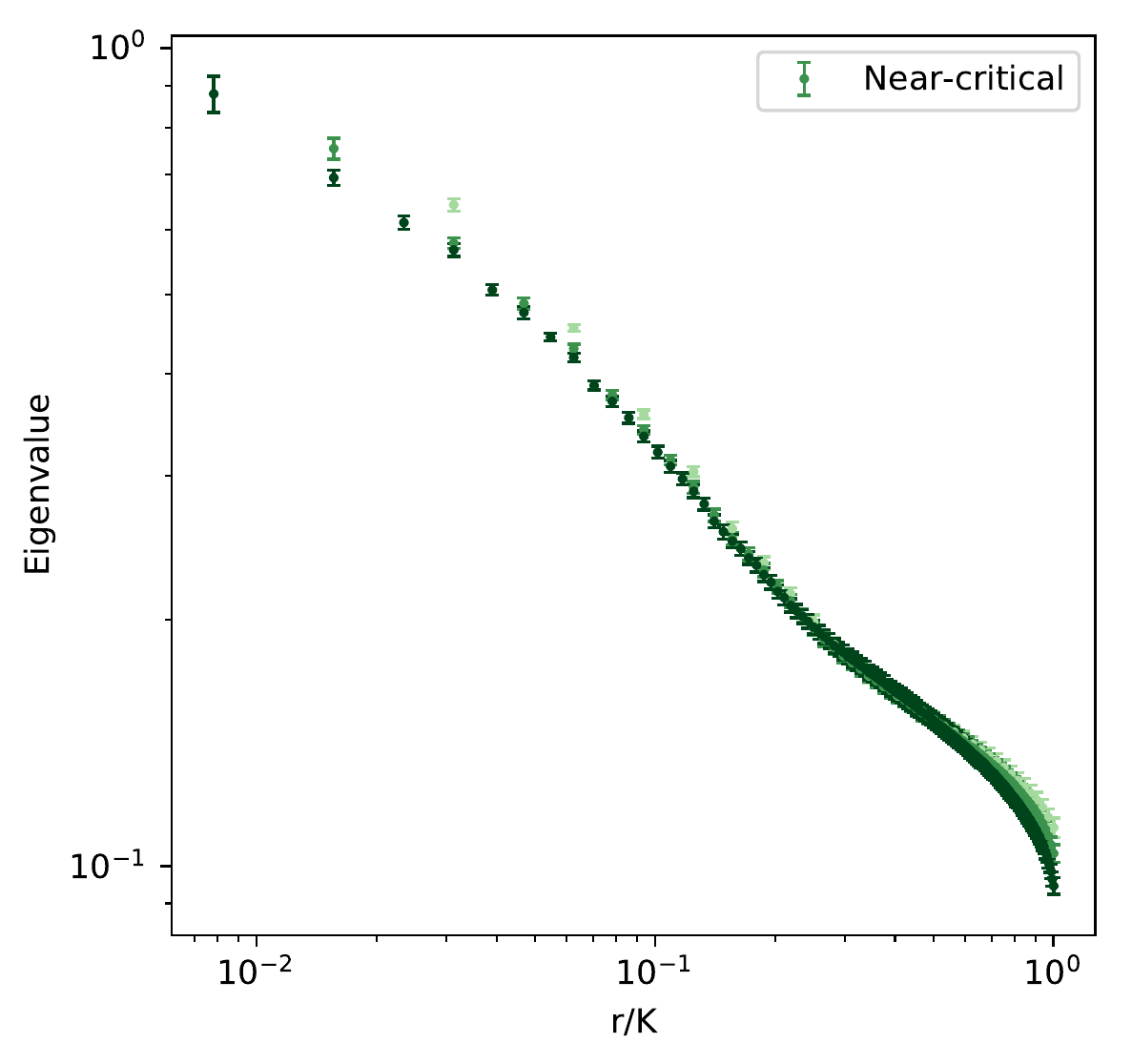}
    \end{minipage}
    \hfill
    \begin{minipage}{0.23\textwidth}
        \centering
        \includegraphics[height=4cm]{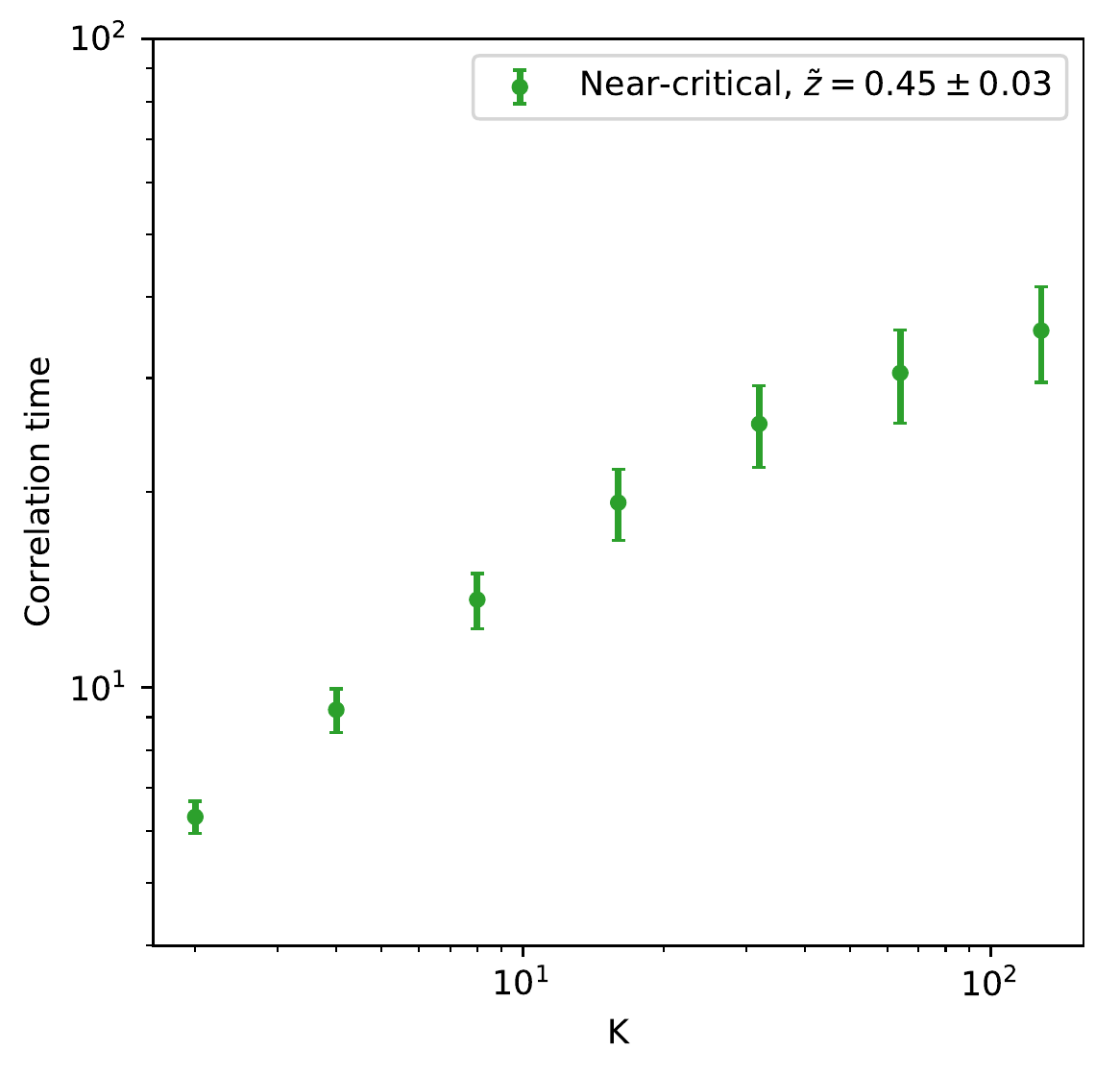}
    \end{minipage}
    \caption{\footnotesize Scaling of the eigenvalues of the covariance matrix for $K = 32, 64, 128$ (from brighter to darker color) and of the autocorrelation times during the coarse-graining for $\lambda \gtrapprox \lambda_c$. The spectrum of the covariance matrix does not show a power law decay, and the scaling of the autocorrelation times is not as convincing as the critical case shown in Figure 3.}
    \label{fig:nc_corr_structure}
\end{figure}

On the other hand, in Figure \ref{fig:nc_corr_structure} the eigenvalues of the covariance matrix do not display an evident power law scaling as we change the cluster size, and the scaling of the autocorrelation time function is not significant, especially for larger clusters.

\begin{figure}[t]
\centering
\begin{minipage}{0.23\textwidth}
        \centering
        \includegraphics[height=4cm]{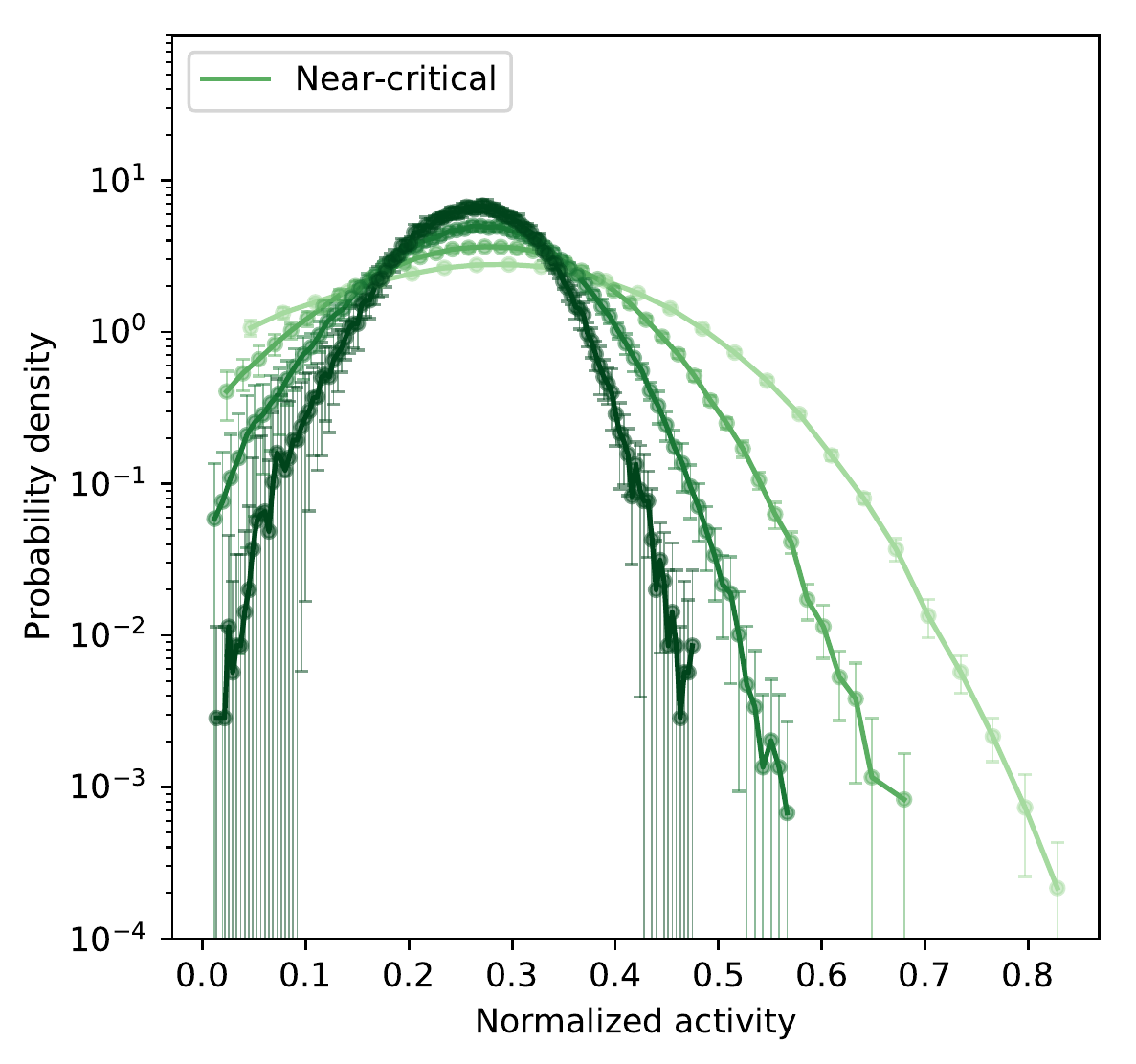}
    \end{minipage}
    \hfill
    \begin{minipage}{0.23\textwidth}
        \centering
        \includegraphics[height=4cm]{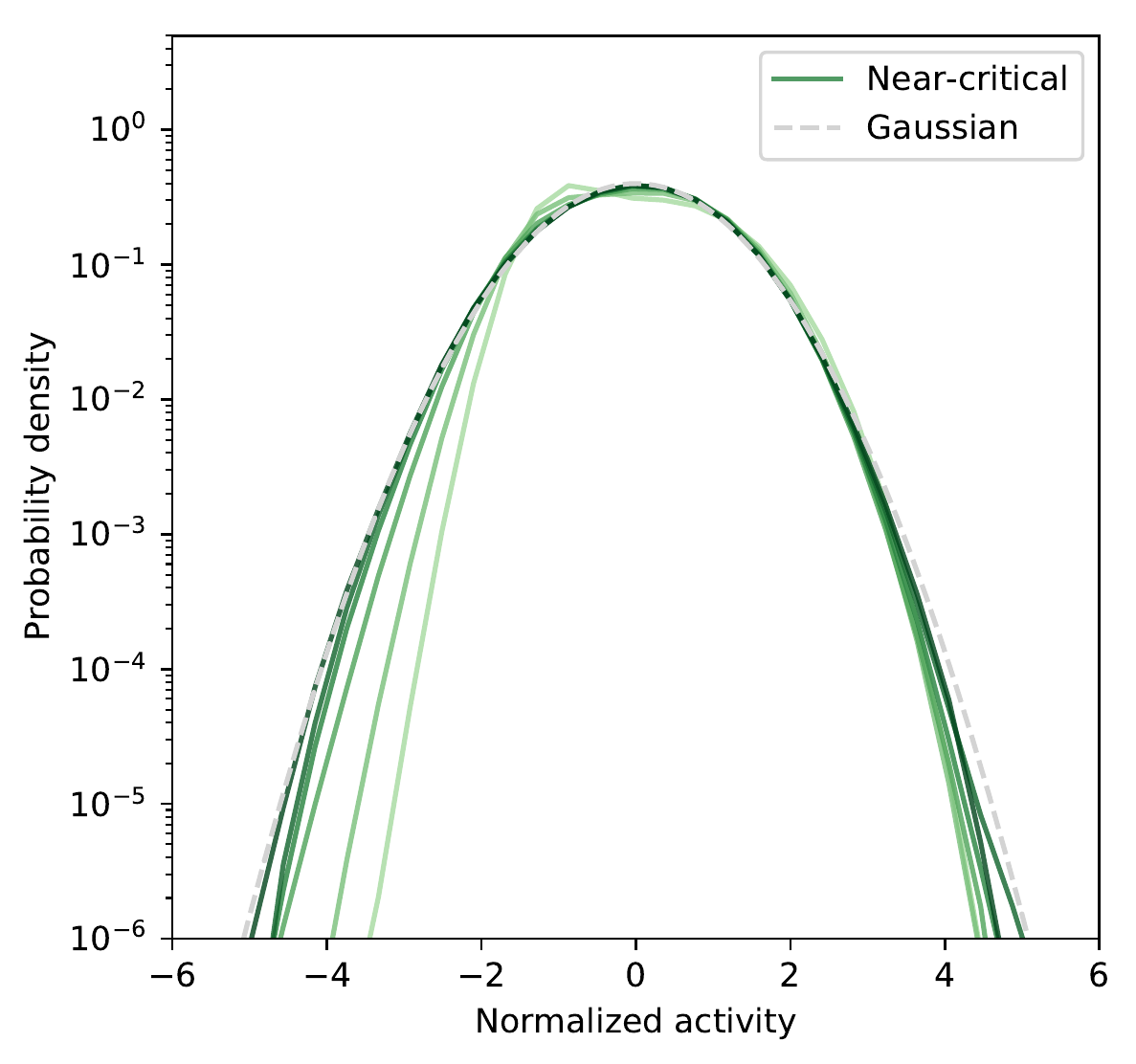}
    \end{minipage}
    \caption{\footnotesize Evolution of the joint probability during the coarse-graining for $\lambda \gtrapprox \lambda_c$. We show the results for $K = 32, \dots, 256$ and for $N/8, \dots, N/128$ modes. Both of them are comparable with the super-critical case, and in particular in momentum space we do not see the non-Gaussian tails typical of criticality.}
    \label{fig:nc_probability}
\end{figure}

The most convincing results to discriminate between critical and quasi-critical state are the joint probability distributions (\ref{eqn:full_probability}, \ref{eqn:cg_momentum_dist}) that we show in Figure \ref{fig:nc_probability}, in particular the one in momentum space. We do not see the non-Gaussian tails that we previously found at the critical point, which is expected since away from criticality the variables are much less correlated with one another and they are eventually dominated by the central limit theorem. 

Nevertheless, following \cite{bib:touboul2017power} in Appendix B we introduce a simple model of conditionally independent neurons that shows a non-trivial form of the joint probability distribution along the coarse-graining in momentum space, although the system is clearly non in a critical state. At the same time, for this model we find that the clustering of maximally correlated pairs of variable seems to work in identifying the model as not critical, because there is no interaction between the neurons themselves.

All these results therefore suggest that in order to infer the system state is crucial not to focus on a single observable, but analyze all the different coarse-grained quantities.

\section{Discussion}
The phenomenological approach introduced in \cite{bib:cg_bialek, bib:bialek_prl} has two considerable advantages: it is model independent so it can be applied directly to the data, and it is stable with respect to the presence of long-range interactions. In this work we tested the PRG both in equilibrium models (see Appendix C for the Ising model), where we expect that it is be able to distinguish between critical and non-critical phases, but also in non-equilibrium models, such as the contact process. We have found that the super and sub-critical regimes can be easily recognized, even though the nature of the phase transition is qualitatively different from the one of the Ising model. 

At the same time we have highlighted that quasi-critical states are difficult to infer, especially if only a subset of physical quantities is analyzed. In non-trivial dynamical models, such as the contact process, the strategy that works best seems to be the one related to the correlation structure. For instance, the presence of non-Gaussian tails in the joint probability distribution of the coarse-grained variables in momentum space, in Figure \ref{fig:probability} and Figure \ref{fig:nc_probability}, might be a good signature of a possible underlying criticality, but at the same time clustering maximally correlated variables fails as we approach the critical point, Figure \ref{fig:var_sp} and Figure \ref{fig:nc_var_sp}. Interestingly, in considerably simpler models (such as the one of Appendix B) the situation is reversed. Hence, in principle, one needs to study both the approach via maximally correlated pairs and in momentum space. Notably, this is the case of \cite{bib:cg_bialek, bib:bialek_prl}, hence the claim of the authors that the neuronal dynamics is critical should still hold.

However, we believe that this approach gives in general a set of necessary conditions for criticality rather than sufficient ones. For instance, the presence of a non-trivial distribution of the coarse-grained variables in momentum space is a necessary condition for criticality because it implies that the underlying variables are strongly correlated, but the convergence in the critical case is hardly clear and calls for particular attention when dealing with experimental data. One might also wonder, inspired by simple non-critical models as the one studied in \cite{bib:touboul2017power}, if there are distinct contributions to the dynamical evolution of the system - an intrinsic one, given by the interaction between the microscopic degrees of freedom, and an extrinsic one, given by some external driver \cite{bib:swain2002intrinsic}. The binomial model we propose in Appendix B is an archetypal example of the latter, and further research is needed if we want to distinguish the two contributions and, eventually, understand which of them contributes to poising the system at criticality.

Overall we believe that, as it is, this PRG might be considered as a better method to infer the presence of a critical state with respect to typical inference methods based on the identification of avalanches in both size and duration with particular exponents \cite{bib:plenz, bib:plasticity, hesse2014self, bib:colloquium, bib:fontenele2019criticality}.

The existence in the data of the signatures of criticality we have highlighted - such as non-Gaussian tails in the distribution of coarse-grained variables in momentum space - are possibly a powerful and stable indicator to characterize the state of a neuronal network. Extending these methods and test them systematically as in the present work might provide further insights in the understanding of the role of criticality in living systems. For example, recent works suggest that the actual transition in the brain dynamics is not between low and high neural activity states, but rather between an asynchronous and synchronous states \cite{bib:disanto2018landau,bib:fontenele2019criticality}. An interesting future direction would be to extend the application of this PRG so to characterize different types of critical transitions in terms of the coarse grained variables.

\section{Acknowledgments}
\noindent S.S. acknowledges UNIPD for STARS 2018 grant.

\vspace{0.5cm}

\noindent \hrulefill


\section*{Appendix A: Spectral properties of the covariance matrix}
\noindent Let us briefly show that a power law spectrum of the covariance matrix is the consequence of the algebraic decay of the spatial correlation function at the critical point. Consider the covariance matrix
\begin{align}
C_{ij} = \ev{\sigma_i\sigma_j} - \ev{\sigma_i}\ev{\sigma_j}.
\end{align}
In a system with translational invariance each element of the covariance matrix is given by $C_{ij}=C(\vb x_i - \vb x_j)$ for some function $C$, whose Fourier transform is given by
\begin{align*}
C(\vb k, \vb q) & = \frac{1}{N}\sum_{i,j} C(\vb x_i - \vb x_j) \, e^{-i\vb x_i \cdot\vb k}e^{-i\vb x_j \cdot \vb q} \\
& = \delta_{\vb k, -\vb q} G(\vb k)
\end{align*}
where
\begin{align*}
G(\vb k) = \sum_n e^{-i\vb x_n \cdot\vb k} C(\vb x_n).
\end{align*}
Hence the covariance matrix has entries given by
\begin{align*}
C(\vb x_i - \vb x_j) = \frac{1}{N} \sum_{\vb k} e^{i\vb k \cdot (\vb x_i - \vb x_j)}G(\vb k),
\end{align*}
which means that in Fourier space the covariance matrix is diagonal. In fact, it is easy to show that the eigenvalues are given by the Fourier transform of the correlation function $G(\vb k)$, since
\begin{align*}
    \sum_{\vb x_j} C(\vb x_i - \vb x_j)e^{i\vb k \cdot \vb x_j} & = e^{i\vb k \cdot \vb x_i} \sum_{\vb x_j} C(\vb x_i - \vb x_j)e^{-i( \vb k \cdot \vb x_i-\vb k \cdot \vb x_j)}\\
    & = e^{i \vb k \cdot \vb x_i} G(\vb k)
\end{align*}
hence $e^{ikx}$ is a eigenfunction of eigenvalue $G(\vb k)$.

This has a non trivial implication for the eigenvalue spectrum of the covariance matrix in a critical system, where we expect the algebraic decay $G(r) \sim r^{-(d-2+\eta)}$. Since the eigenvalues are the Fourier transform of the correlation function, we shall write
\begin{align*}
\lambda_{\vb k} & \sim \int d^dr \, e^{i\vb k \cdot \vb r} r^{-(d-2+\eta)} \sim \frac{1}{|\vb k|^{2-\eta}}.
\end{align*}
If this is a decreasing function of $|\vb k|$, that is if $\eta<2$, then we consider a ranking of eigenvalues from small momentum to large momentum. Hence the highest eigenvalue has rank $r=1$, which implies
\begin{align*}
r[\lambda_{\vb k}] & = \sum_{\vb k'} \mathbb{I}[\lambda_{\vb k'}> \lambda_{\vb k}] = \sum_{\vb k'}\mathbb{I}[|\vb k'|<|\vb k|] \\
& \approx L^d \int d^dk'\,\theta(|\vb k'| < |\vb k|) \\
& \sim (L|\vb k|)^d.
\end{align*}
This implies that the eigenvalues of the covariance matrix decay as a power law of their rank, namely
\begin{align}
\label{eqn:spectrum_covariance}
\lambda_r \sim \frac{1}{r^\mu}
\end{align}
where $\lambda_1 \ge \lambda_2 \ge \dots \lambda_N$ and $\mu = (2-\eta)/d$.


\section*{Appendix B: Model of conditionally independent neurons}
\noindent The convergence of the joint probability distribution in Equation (\ref{eqn:cg_momentum_dist}) is related to the spectrum of the covariance matrix, but one should be careful when considering its relation with criticality. Consider $N$ random variables $\left(\sigma_1^{t}, \dots, \sigma_N^{t}\right)$. At each time $t$ the distribution of the $i$-th variable, which we can think of as a neuron that can be either active or inactive, is a simple binomial distribution with parameter $\xi_i(t)$. However, we consider the case in which also $\bm{\xi}(t) = (\xi_1(t), \, \dots, \, \xi_N(t))$ is itself a random variable distributed according to some distribution $p(\bm{\xi})$, so that the $N$ neurons are conditionally independent. We can think of this case as the one of neurons that follow a common external dynamics, represented by $p(\bm{\xi})$, but otherwise show no intrinsic dynamical features.

The probability that a neuron is either active or inactive is then a binomial distribution conditioned to the value of $\xi_i(t)$, that is
\begin{equation*}
    p(\sigma_i^t \,|\, \bm\xi = \bm\xi(t)) = 
    \begin{cases}
        \xi_i(t) & \sigma_i^t = 1 \\
        1 - \xi_i(t) & \sigma_i^t = 0
    \end{cases},
\end{equation*}
and the each neuron is described by the joint probability $p(\sigma_i^t, \bm\xi) = p(\sigma_i^t \,|\, \bm\xi = \bm\xi(t))\, p(\bm{\xi})$.

Since the coarse-graining procedure depends on the equal-time covariance of the neurons $\text{cov}(\sigma_i, \sigma_j) = C_{ij}$, we need the marginal probabilities
\begin{equation*}
    p(\sigma_i^t) = 
    \begin{cases}
    \displaystyle\int d\xi^*_i \xi^*_i p(\xi^*_i) = \ev{\xi_i} & \sigma_i^t = 1\\
    \displaystyle\int d\xi^*_i (1-\xi^*_i) p(\xi^*_i) = 1-\ev{\xi_i} & \sigma_i^t = 0
    \end{cases}
\end{equation*}
and
\begin{equation*}
    p(\sigma_i^t, \sigma_j^t) = 
    \begin{cases}
    \ev{\xi_i\xi_j} & \sigma_i^t = 1, \sigma_j^t = 1\\
    \ev{\xi_i}-\ev{\xi_i\xi_j} & \sigma_i^t = 1, \sigma_j^t = 0\\
    \ev{\xi_j}-\ev{\xi_i\xi_j} & \sigma_i^t = 0, \sigma_j^t = 1\\
    1 + \ev{\xi_i\xi_j}-\ev{\xi_i}-\ev{\xi_j} & \sigma_i^t = 0, \sigma_j^t = 0\\
    \end{cases}.
\end{equation*}
It is then trivial to see that, even if the neurons are not correlated, their covariance is not vanishing but depends on the covariance of $p(\bm \xi)$,
\begin{equation*}
    C_{ij} =
    \begin{cases}
        \ev{\xi_i\xi_j} - \ev{\xi_i}\ev{\xi_j} & i \ne j \\
        \ev{\xi_i}(1-\ev{\xi_i}) & i = j
    \end{cases}.
\end{equation*}

Let us consider the simple case of $\xi_i = \xi_j$ $\forall i, j$, so that at each time all the neurons fire with the same probability $\xi$, and take $p(\xi_i)$ to be a uniform distribution. In this case the covariance matrix is simply
\begin{equation*}
    C_{ij} = a \delta_{ij} + b (1-\delta_{ij})
\end{equation*}
with $a = 1/4$ and $b = 1/12$. The eigenvalues of this matrix are given by
\begin{align*}
    & \lambda_1 = a+(N-1)b \quad m = 1 \\
    & \lambda_2 = a-b \qquad\qquad\,\,\, m = N-1
\end{align*}
where $m$ is the corresponding multiplicity. Therefore, there are $N-1$ eigenvalues with the same value.

The eigenvector associated to the highest eigenvalue is $1/\sqrt{N} (1, \dots, 1)^T$, but there is no obvious choice for the other eigenvectors in Equation (\ref{eqn:projector}) because the ranking is ill-defined. However, from a numerical standpoint the spectrum of the covariance matrix will not be degenerate, so if we simulate the model we can try to apply the procedure regardless.

\begin{figure}[t]
\centering
\begin{minipage}{0.23\textwidth}
        \centering
        \includegraphics[height=4cm]{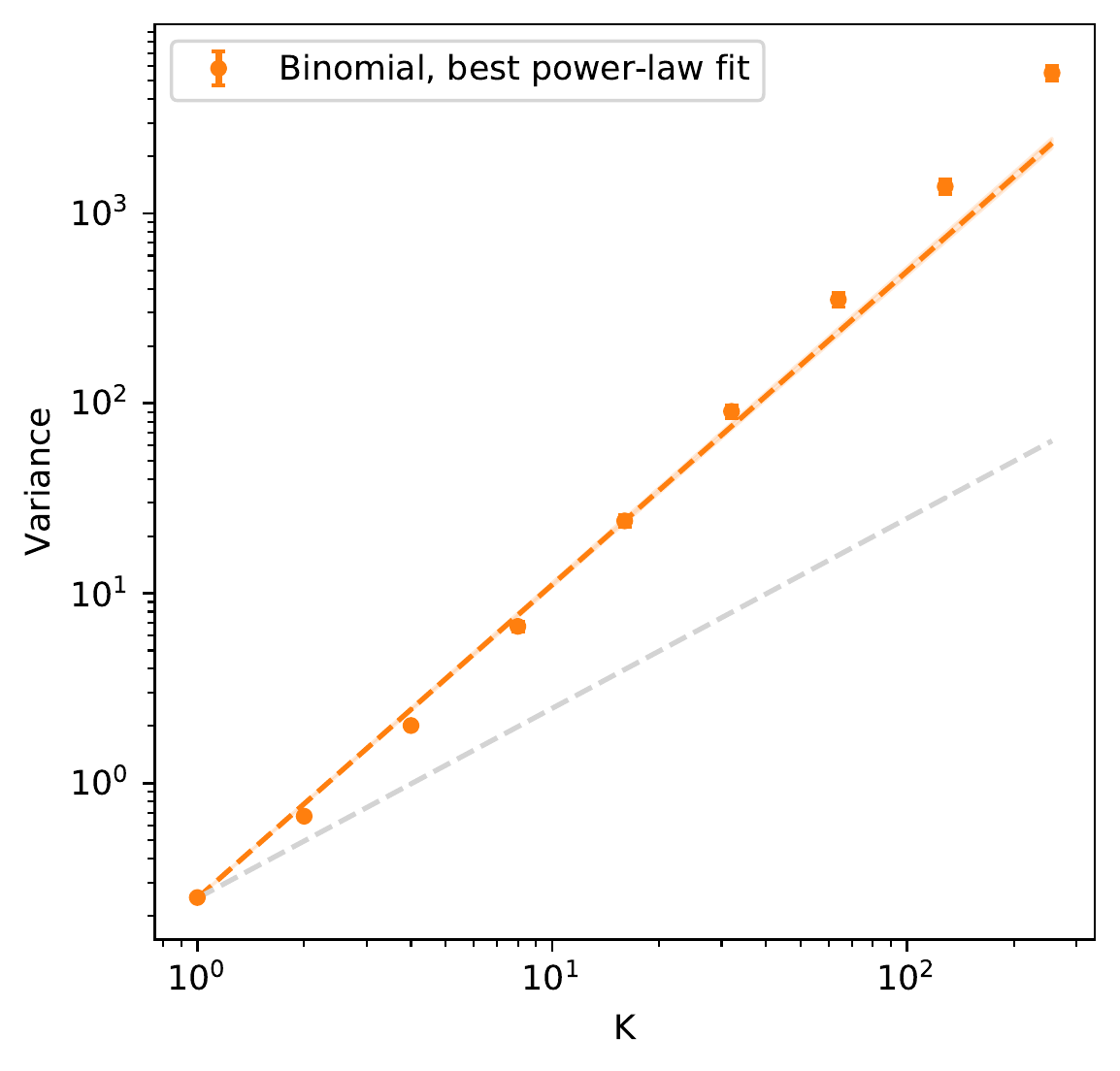}
    \end{minipage}
    \hfill
    \begin{minipage}{0.23\textwidth}
        \centering
        \includegraphics[height=4cm]{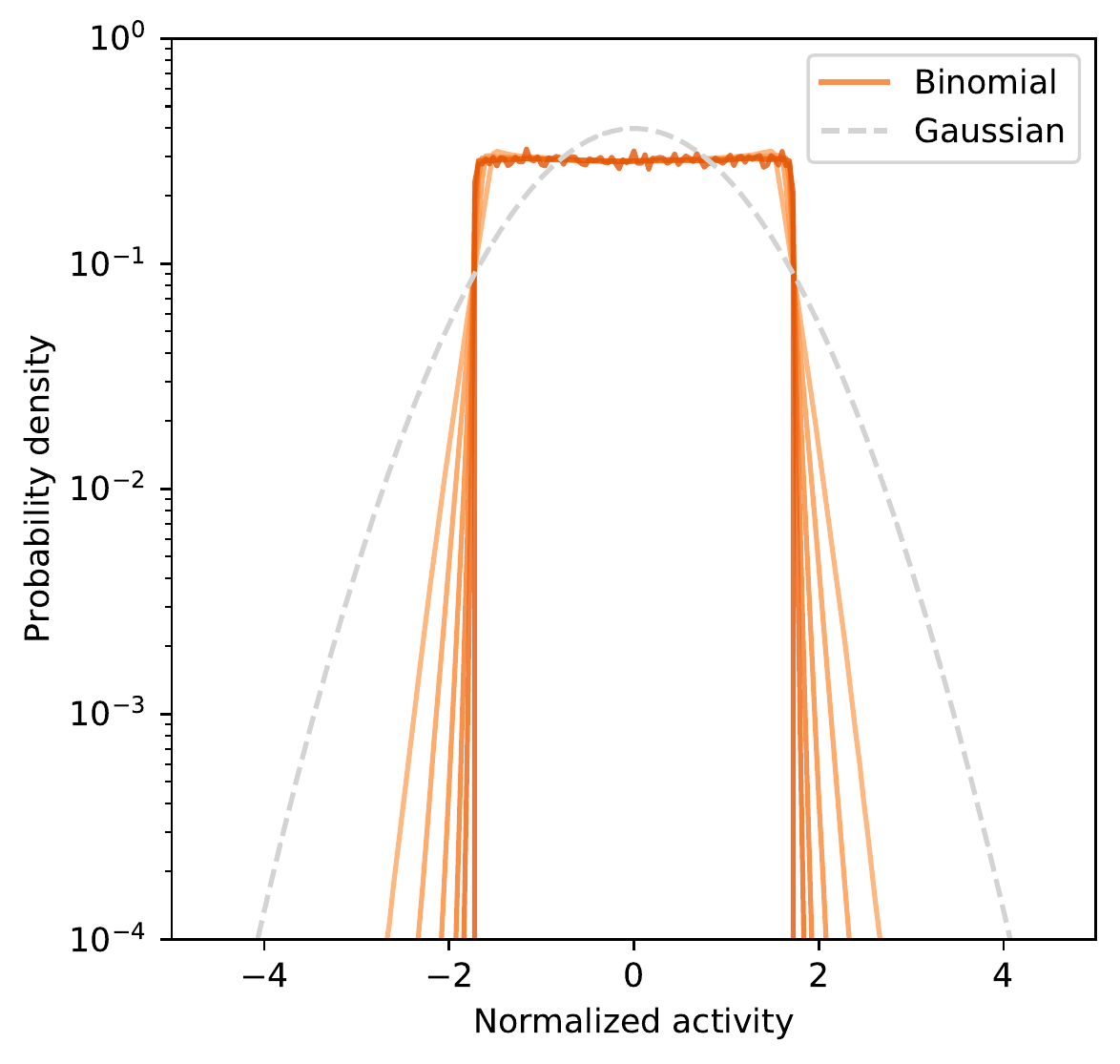}
    \end{minipage}
    \caption{\footnotesize Failure of scaling of the variance in the binomial model and the non-Gaussian distribution resulting from the procedure in momentum space.}
    \label{fig:binomial}
\end{figure}

As we can see from Figure \ref{fig:binomial} the joint probability does not converge to a Gaussian, even though there is nothing critical about the underlying dynamics. Hence the proposed coarse-graining in momentum space fails for a simple set of conditionally independent binomial variables, albeit it seemed to be the most promising procedure for the supercritical contact process in the vicinity of the critical point. On the other hand, and perhaps not surprisingly, in this model the proposed coarse-graining procedure via maximally correlated variables does work: in fact, since the off-diagonal elements of the covariance matrix are all equal we are randomly pairing neurons together and no scaling property emerges.

This simple model shows once more how careful one should be when employing these kind of procedures. All in all, the two approaches combined seem to work well, in the sense that a system might really be critical if both of them indicate the presence of an underlying scale invariance. However, when we take them individually they might point in the wrong direction.

\section*{Appendix C: Ising Model}
\noindent We also test the coarse-graining procedure in the $2D$ Ising model. We do not show the results explicitly for the sake of brevity, but they are in line with what one might expect.

In the disordered phase at $T>T_c$ the coarse-graining drives the system towards a behavior that is comparable with the one of independent random variables, very much like one would expect from a usual block-spin transformation in real space.

For $T<T_c$, instead, the behavior of coarse-grained variables resembles the one of a perfectly ordered system.

However, as we lower the temperature we see a non-trivial effect due to the spontaneous symmetry breaking that occurs at the transition. In fact, the Ising model in its ordered phase is essentially low dimensional \citep{bib:pca_ising} in the sense that one single eigenvalue eventually dominates the spectrum of the covariance matrix. Once we take this into account, we find that at criticality its spectrum scales with an exponent $\mu = 0.88 \pm 0.03$ which is perfectly compatible with the value $\mu = 7/8$ one gets from the exact solution. Hence, in this case of an equilibrium phase transition with spontaneous symmetry breaking, this procedure does identify two distinct phases \footnote{One should note that since the Ising variables are $\sigma_i = \pm 1$, we cannot define in the same way a silence probability and the corresponding free energy (\ref{eqn:cg_free_energy})}.

\end{document}